# Resonance-enhanced waveguide-coupled silicon-germanium detector

L. Alloatti[a),*], R. J. Ram

*Massachusetts Institute of Technology, Cambridge, MA 02139, USA*
[a)]*Current address: Institute of Electromagnetic Fields (IEF), ETH Zurich, Zurich, Switzerland*
[*]*Electronic mail: luca.alloatti@gmail.com*

A photodiode with 0.55±0.1 A/W responsivity at a wavelength of 1176.9 nm has been fabricated in a 45 nm microelectronics silicon-on-insulator foundry process. The resonant waveguide photodetector exploits carrier generation in silicon-germanium (SiGe) within a microring which is compatible with high-performance electronics. A 3 dB bandwidth of 5 GHz at -4 V bias is obtained with a dark current of less than 20 pA.

Single-chip microprocessors can exceed a compute capacity of five trillion floating-point operations per second (5 TFLOPS)[1] therefore requiring an input/output (I/O) bandwidth of 40 Tb/s consistent with the approximately "one byte I/O per flop" rule-of-thumb.[2,3] However, the physical limitations of electrical interconnects –which are constrained by RF losses, electromagnetic interference, power dissipation, and package pin density– typically limit the available bandwidth to a tenth of the peak bandwidth required.

Monolithic integration of optical transceivers side-by-side with billion-transistor circuits has the potential to overcome these limitations. However, achieving the necessary transistors' yield together with high photonics performance has been a major challenge. Monolithic approaches developed so far have followed the path of modifying existing electronic processes by adding fabrication steps and materials such as pure germanium for photocarrier generation[4,5] with the risk of shifting the transistor specifications and decreasing the fabrication yield. These processes moreover exploit 90 nm or older nodes which are not currently utilized for building high-performance computers (HPC).[6,7]

An alternative approach consists of designing photonic components in existing CMOS nodes without violating any design rule and without requiring any modifications to the process flow –so-called "zero-change CMOS".[8,9] Within the GlobalFoundries (formerly IBM) 45 nm 12SOI node we have recently demonstrated a complete zero-change photonic toolbox comprising waveguides with 5 dB/cm,[8] grating-couplers,[10] 5 Gbps modulators,[11] and 32 GHz photodetectors.[7] These components enabled the first realization of an optical link between a microprocessor and an external memory.[12] However, the responsivity of the first photodiodes was limited to ~0.02 A/W and directly impacted the power efficiency of the link.[12]

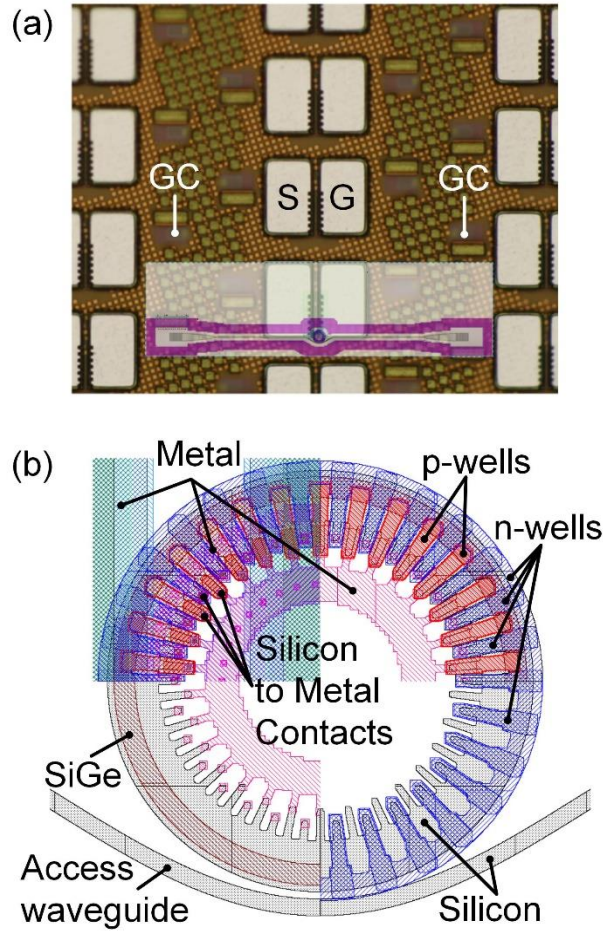

Fig. 1. Photodetector geometry. (a) Microphotograph of the fabricated chip showing a sweep of devices. The ground-signal (GS) electrodes can be easily recognized along with the grating-couplers (GC) apertures through the metal stack. The layout of a single device is superimposed. (b) Layout of the cavity with the access waveguide. The cavity region is divided in four quadrants with different mask layers highlighted/hidden. In the top-left quadrant most of the layers are activated including the high-doping regions (dark red and dark blue), the inner and outer metal contacts and some vias. The top-right quadrant highlights the relative positions of the n-type and p-type well implants. In the bottom-left quadrant only the 500 nm wide SiGe ring is visible together with the silicon cavity and one metal layer. In the bottom-right quadrant only the n-implants and the silicon cavity are highlighted. The T shape of the implants is recognizable.

In this work, we present a zero-change CMOS photodiode with a responsivity of 0.55±0.1 A/W –a twenty-fold increase over previous results.[12] Similar to our first demonstration, the photodiode exploits carrier generation in silicon-germanium (SiGe) which is already present in the 45 nm 12SOI node for stressing the channels of p-FETs.[7] To increase the responsivity, here we exploit resonance-enhancement in a microring. The disk-like cavity supports a whispering gallery mode for effectively separating the optical field from the metal contacts.[11] The cavity has a radius of 5 $\mu$m and is etched into the crystalline silicon which is normally used in the 45 nm process for realizing the body of the transistors.[8] This type of cavity has already been exploited for building modulators,[11] but the gap between the bus waveguide and the ring has been reduced for compensating the higher absorption of the detector, low-resistance contacts have been added between silicon and vias, and the number of pn-junctions has been reduced to minimize the device resistance and capacitance. Inside this cavity, a 500 nm wide ring is partially etched into the silicon for forming the heteroepitaxially-grown SiGe absorption region, Fig. 1. The SiGe alloy has a germanium content between 25% and 35% based on literature data and on previous experiments.[7] Two n-type and two p-type spoke-shaped well-implants form interdigitated pn-junctions in the SiGe region and connect the active region to the contacts located on the inner radius, Fig. 1(b). The p-type and n-type spokes have different lengths so that a part of the pn-junction is located in the middle of the SiGe section. These well-implants are performed prior to SiGe deposition and affect the silicon only. It is not known whether the SiGe has been doped chemically during the deposition, and it is possible that it is nearly intrinsic. The implant spokes are narrower in the inner part of the cavity (T-shapes) for reducing the parasitic junction capacitance between the SiGe region and the electrical contacts. Source-drain (S/D) well implants, halo/extension implants and silicidation complete the electrical contact to the high-frequency ground-signal (GS) electrodes.[7]

To facilitate testing, broadband grating couplers (1170 nm -1560 nm) are used to couple light in and out of the waveguide and cause a wavelength-dependent loss of at least 10 dB each, although optimized grating couplers have been demonstrated elsewhere on this chip with 1.2 dB insertion loss.[10] The waveguides have been designed by a fully-scripted photonic-design automation (PDA) tool based on Cadence with automatic DRC-cleaning and layer generation.[9] The present device belongs to a sweep of 60 variations obtained by permuting 5 different coupling gaps, 6 doping patterns and 2 SiGe ring widths. The chip was taped-out in March 2015 and manufacturing was completed by September 2015. Among the variations on the doping patterns the present device shows the highest bandwidth. Other variations contain T-shaped spokes having equal dimensions for both the n-implants and p-implants extending from the inner radius to the outer radius of the cavity so that the junctions cut the SiGe region radially. No halo nor extension implants (which would affect the SiGe as well) have been used on the optical cavities yet, and they may be used to improve the junction doping profile in future generations.

The current-voltage characteristic is shown under different illumination conditions in Fig. 1(a). The dark current is smaller than 20 pA in the reverse bias range -5 V to 0 V corresponding to a dark current density of 0.116 mA/cm$^2$. This value is significantly smaller than for the majority of germanium-based photodiodes demonstrated so far in silicon photonics,[13,14,15,16] and is close to the thermionic emission limit of germanium photodiodes which is ~$10^{-2}$ mA/cm$^2$.[16] The small dark current is attributed to a low density of defects and dislocations.

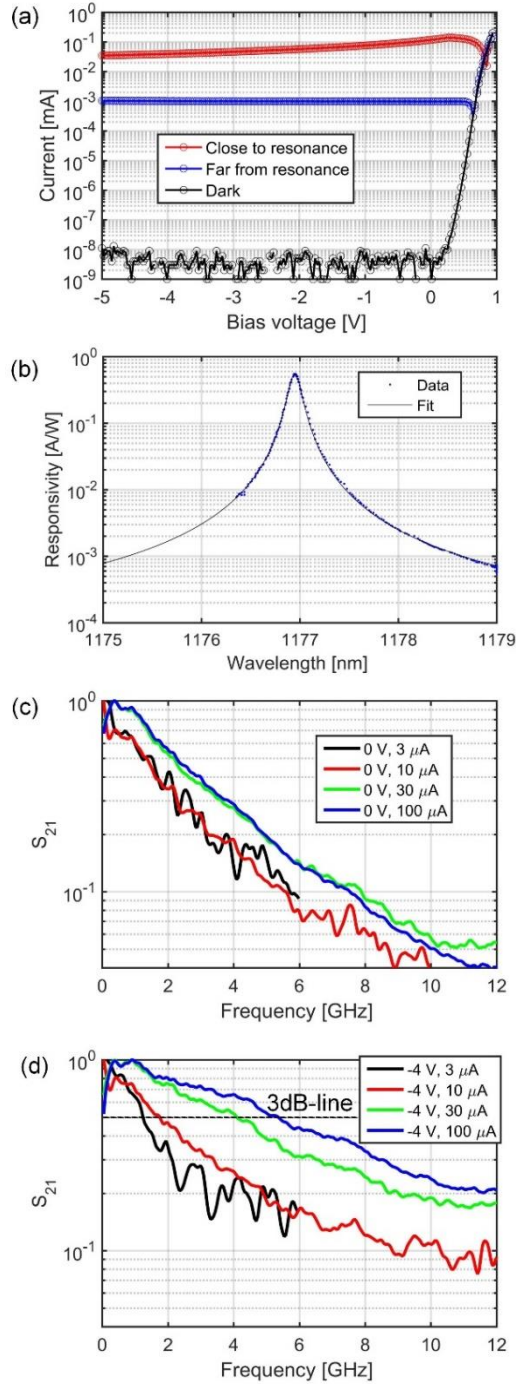

Fig. 2. Device performance. (a) Current-voltage characteristics for different illumination conditions. The dark current is smaller than 20 pA for -5 V to 0 V bias. When the wavelength is far from resonance the reverse-bias photocurrent is approximately constant. The variation of the photocurrent when the wavelength is close to resonance is attributed to self-heating effects which detune the cavity. (b) Responsivity vs. wavelength. A maximum of 0.55 A/W is observed. Data was recorded at low optical powers first by increasing the wavelength and then by decreasing it for excluding heating-induced drifts. The thermally-tuned laser source (QDLaser model QLD1161-8030) was limited to wavelengths larger than ~1176.4 nm. (c) Frequency response at 0 V bias for different in-cavity optical powers (different average currents). In the low-frequency limit the VNA measurements were not reproducible so that a 30% error should be taken into account on the normalization. (d) Frequency response at -4 V bias for different average currents. A 3 dB bandwidth of 5 GHz is obtained at -4 V bias for an average current of 100 $\mu$A.

The responsivity vs. wavelength is shown in Fig. 2(b) and reaches a peak value of 0.55±0.1 A/W at ~1176.9 nm corresponding to a quantum efficiency of 58%. The data has been fitted with the model of a ring resonator having a FWHM transmission $\delta\lambda_{FWHM}$ = 0.138 nm (29.9 GHz) corresponding to a loaded Q = $\lambda/\delta\lambda_{FWHM}$ = 8530. The responsivity was measured by recording photocurrent and optical transmission at the same time. The current was measured at low optical powers to minimize the drift of the resonant frequency by self-heating,[17,18] and the insertion loss of the input and output grating couplers was subtracted out. The measurement was repeated a second time with exchanged input and output fiber connectors such as to verify that the input and output grating couplers caused the same optical loss.[7] The free spectral range (FSR) is 13.75 nm and the transmission at resonance is about -15 dB. The excitation of a higher order mode causes an additional responsivity peak red-shifted by ~5 nm from the main resonance and suppressed by about 19 dB.

The bandwidth of the device was measured by contacting the GS electrodes with a 50 $\mu$m pitch GS probe of Cascade Microtech (model Infinity I67-A-GS-50). The reference plane was set at the V-connector of the probe. The frequency response was measured with a 40 GHz VNA (HP8722D) and the frequency-response of the setup (comprising modulator, RF cables and bias-T) was calibrated with a reference photodiode (Discovery Semiconductors, model DSC30-3-2010) of known frequency-response.[7] The frequency response was recorded for different bias voltages and different offsets from the resonant wavelength therefore resulting in different average currents (the in-waveguide power was kept constant), Fig. 2. For 0 V bias the frequency response is almost independent of the optical power in the waveguide and follows a nearly exponential roll-off.[19] For -4 V bias we observe that higher in-cavity optical powers (larger average currents) correspond to higher bandwidths. For both bias voltages the frequency roll-off suggests that the device is not limited by the RC time constant. For -4 V bias, the 3 dB bandwidth varies between ~2 GHz and 5 GHz when the average current changes between 10 $\mu$A and 100 $\mu$A. Since the frequency response was found earlier to be very sensitive to the position of the pn-junction relatively to the SiGe region[7] we expect that faster devices can be obtained by optimizing the doping profile. Eye diagrams have been recorded at 5 Gb/s and 12.5 Gb/s (PRBS length $2^{31}$-1) with an Agilent waveform analyzer (model 86108A with 50 ohm termination), Fig. 3. In these experiments no optical nor electrical amplifiers have been used. Open eyes are obtained at 5 Gb/s with -1 V bias and at 12.5 Gb/s at -4 V bias with 120 $\mu$A average current.

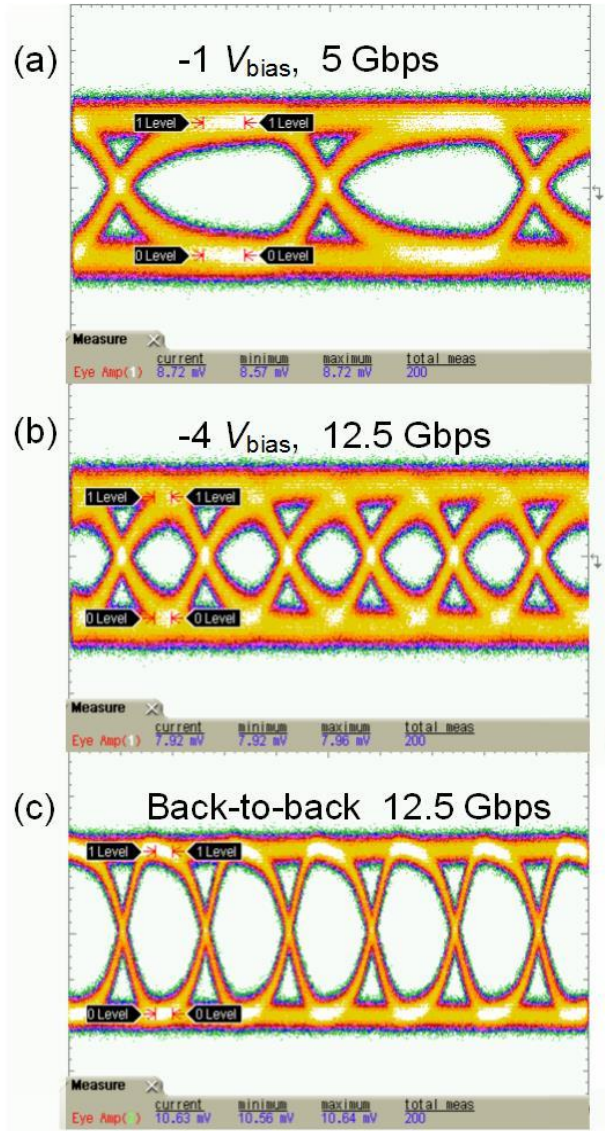

Fig. 3. Eye diagrams. (a) 5 Gb/s and -1 V bias. (b) 12.5 Gb/s and -4 V bias. (c) Back-to-back reference at 12.5 Gb/s taken with a 30 GHz commercial photodiode. All the eye diagrams have a vertical scale of 3 mV/div. The same number of measurements was taken for all eye diagrams.

In conclusion, we have demonstrated a photodiode with a responsivity of 0.55±0.1 A/W at 0 V bias in zero-chance CMOS. The photodiode is realized in the 45 nm 12SOI CMOS node, which is widely used in high-performance computers (HPC). The silicon-germanium used for optical absorption is grown heteroepitaxially in circular silicon pockets and has a low germanium content in contrast to mainstream silicon-photonic photodiodes. We are not aware of previous demonstrations of germanium or silicon-germanium having boundaries which are not aligned with a major crystalline axis. High-responsivity photodiodes are key components for minimizing the optical power budget in future chip-to-chip transceivers.[12] This detector, when coupled with previously demonstrated receivers having an average sensitivity of less than 3 $\mu$A at 2.5 Gb/s,[17] would correspond to a sensitivity of -22.6 dBm, a 17 dB improvement over previous demonstrations.[12] The resonant design enables the detection of wavelength-division multiplexing (WDM) signals without introducing additional filters.

We acknowledge support by DARPA POEM under award HR0011-11-C-0100 and contract HR0011-11-9-0009. The views expressed are those of the authors and do not reflect the official policy or position of the DoD or the U.S. Government. We thank Amir Atabaki for performing the chip substrate transfer.